\documentclass[]{spie}  

\usepackage{amsmath,amsfonts,amssymb}
\usepackage{graphicx}
\usepackage[colorlinks=true, allcolors=blue]{hyperref}
\usepackage{lipsum}

\title{Versatile Optical Ground Station for Satellite-based Quantum Key Distribution in Abu Dhabi}

\author[1]{Sana Amairi-Pyka}
\author[2]{Christoph Fischer}
\author[1]{Konstantin Kravtsov}
\author[1]{Gianluca De Santis}
\author[1]{Alessandro Grosso}
\author[2]{Edgar Fischer}
\author[2]{Klaus Kudielka}
\author[1]{James A. Grieve}
\affil[1]{Quantum Research Centre, Technology Innovation Institute, 9639 Abu Dhabi, United Arab Emirates}
\affil[2]{GA-Synopta, Eggersriet, Switzerland}

\authorinfo{E-mail: sana.pyka@tii.ae}

\begin{document} 
\maketitle

\begin{abstract}
With the growing number of satellite-based Quantum Key Distribution (QKD) payload launches, it becomes essential to ensure compatibility across different platforms for satellite tracking and quantum signal acquisition. In this paper, the Technology Innovation Institute (TII) presents the development of the Abu Dhabi Quantum Optical Ground Station (ADQOGS) for secure free-space optical communications. With the know-how of GA-Synopta’s experienced engineering team, we have developed a versatile multi-wavelength quantum acquisition and tracking system tailored to support various upcoming space-based QKD missions, crucial for the practical implementation of global quantum communication networks. This system is capable of handling multiple wavelengths, ranging from 600 nm to 1560 nm for downlink beacons and 1530 nm to 1610 nm for uplink beacons. It includes a free-space quantum module adequate to detect QKD signals at 780±10 nm and 850±3 nm and offers spatial and spectral filtering capabilities along with a motorized polarization correction system.  
\end{abstract}


\section{INTRODUCTION}
Quantum key distribution (QKD) is a technology which enables the transmission of cryptographic key material, in which protocol security is based on the laws of quantum mechanics. The technology is range limited by optical losses, motivating the construction of space-based architectures to bridge long distances. In this approach, a QKD satellite may function as a flying trusted node, connecting distant ground stations. However, the reliance on satellites as trusted nodes raises significant security concerns, particularly regarding the ownership and control of cryptographic keys. In fibre-based quantum networks, researchers have developed strategies for routing quantum keys across multiple paths to protect against compromised nodes \cite{zhou2022quantum}. Our design addresses this challenge by supporting redundant satellite key routing across multiple satellite providers, enabling a parallel trusted node approach to quantum key distribution \cite{de2024parallel}. This article presents a new modular design approach for a versatile quantum optical ground station for satellite-based Quantum Key Distribution (QKD). The Abu Dhabi Quantum Optical Ground Station (ADQOGS) is equipped with an advanced Quantum Acquisition and Tracking System, including a motorized QKD receiver module compatible with the major announced QKD satellites. The station’s telescope design enables fast switching between receiver modules, allowing a wide range of satellites to be served. This method enhances the security of QKD over satellite by reducing the risk of one-point trust dependencies.

\section{Versatile acquisition and tracking Systems}
\label{sec:adqogs} 
\begin{figure}
    \centering
    \includegraphics[width=0.9\textwidth]{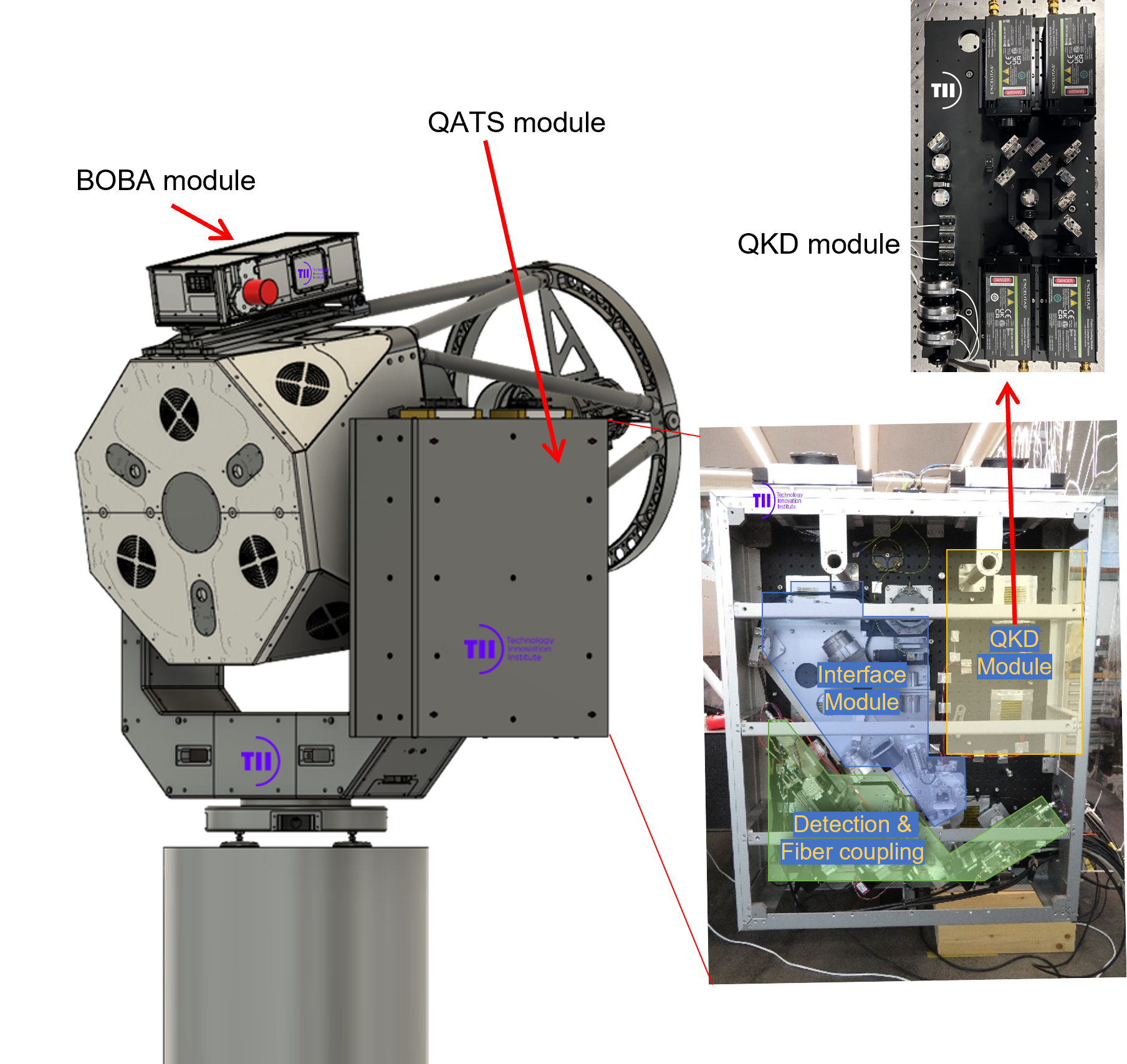}
    \caption{Construction model of ADQOGS's telescope and its sub-systems. QATS: Quantum Acquisition and Tracking System. BOBA: Beacon Optical Bench Assembly. QKD module: free-space receiver module for polarization-encoded Quantum Key Distribution.}
    \label{fig:fullsetup}
\end{figure}

The ADQOGS is located in Al Wathba, Abu Dhabi at $24^\circ 11'$~N, $54^\circ 41'$~E at a relatively low altitude of about 70 m above sea level. The location has a desert climate and an urban-like light pollution level. The dome is equipped with the Ritchey-Chrétien telescope with a main mirror aperture diameter of 80 cm and a focal ratio of f6.85.
The telescope has two interoperable Nasmyth output ports, and one of them will be used for mounting the Quantum Acquisition and Tracking System (QATS) as the multi-modular receiver (Rx), and downlink receiver for SWIR optical communication channels as shown in Figure~\ref{fig:fullsetup}. The multi-band Beacon Optical Bench Assembly (BOBA) will be used to transmit the beacon and as an uplink transmitter (Tx) for free-space optical communication channels. The second Nasmyth port can be used for other applications, i.e. future versions of the receiver. 

\subsection{Multi-Modular QATS design}
The QATS design was inspired by GA-Synopta's earlier work on modular adaptive optics solution \cite{fischer2021modular}. TII team came up with the idea to modify the design so that it becomes adapted to multiple wavelength functionality ranging from 600 nm to 1565 nm for downlink beacons. A multimode fibre output port for 1530-1565 nm amplitude-modulated communication reception is also included. A multimode fibre output port for 1530-1565 nm amplitude-modulated communication reception is also included. TII team added a motorised polarization-based QKD receiver module with 4 free space single-photon detectors, which will be integrated within the QATS box. GA-Synopta team was consulted to simulate, engineer and build the classical section of the QATS box to make the modules more compact and exchangeable. This will allow future development on the same optical chassis supporting more wavelengths and diverse optical communication missions.

\begin{figure}
    \centering
    \includegraphics[width=1\textwidth]{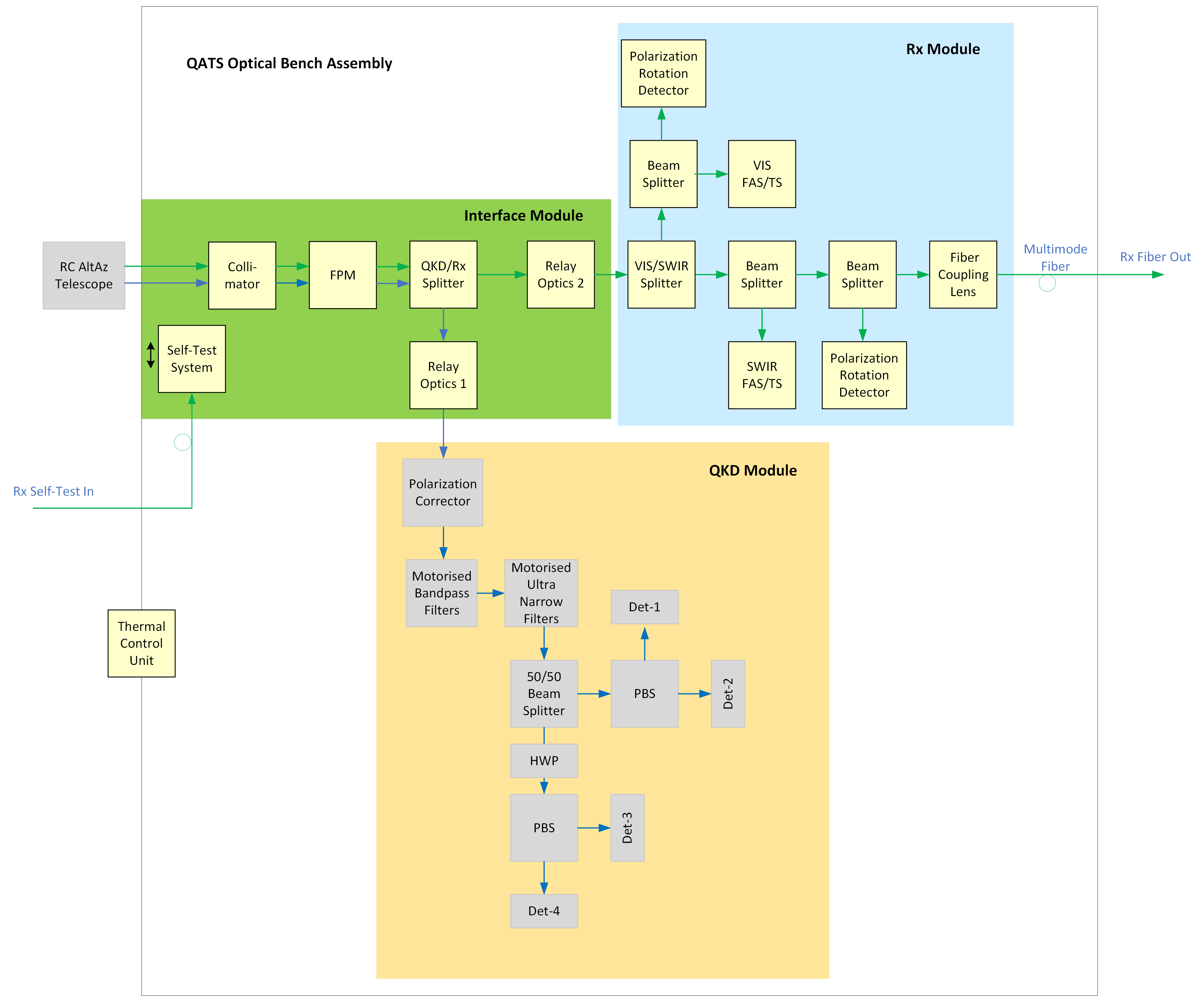}
    \caption{Simplified block diagram of the Quantum Acquisition and Tracking System (QATS). Rx: Receiver path. FAS: Fine Acquisition system. TS: Tracking System. FPM: Fine Pointing Mirror. PBS: Polarised Beam Splitter. HWP: Half Wave Plate. Det: Detector. QKD: Quantum Key Distribution.}
    \label{fig:blockdiagram}
\end{figure}

The QATS is a multi-wavelength tip/tilt stabilized optical receiver system for QKD signals and associated classical downlink beacons. The box is composed of 3 main modules. Figure~\ref{fig:blockdiagram} shows the block diagram of the QATS box.

At the entrance of the QATS box, in the section common to all optical paths, a collimator images the entrance aperture of the telescope into a pupil. A Fine Pointing Mirror (FPM) is placed in this pupil to correctly perform the task of angle-of-arrival compensation (i.e. line-of-sight stabilization). 

The spectral separation between the QKD and the downlink beacon channels occurs right after the FPM. This helps to minimize the number of optical components in the QKD channel path. A dichroic beamsplitter implemented in collimated beam space is best suited for this purpose. In the optical path leading to the QKD receiver, only a small field of view (FoV) is necessary. The FoV has been designed to be 1.5 times larger than the maximum expected size of the seeing disk under the worst conditions for the QKD channel.

No beam splitter can maintain a constant splitting ratio across the entire wavelength range of 600 nm to 1570 nm in the downlink beacon path. Therefore, it is necessary to use a dichroic mirror to spectrally separate the two downlink beacon bands into a VIS/NIR band and a SWIR band.

In both classical receiver channels of the QATS box, a fraction of the incoming downlink light is directed to a fine acquisition and tracking sensor. These detectors, working in a fast control loop, stabilize the telescope's line of sight in conjunction with the FPM. The fine pointer can be off-loaded by sending offset commands to the telescope motion controller.

The effective rotation angle of the polarisation base can be determined by the ground terminal in cases where the space terminals transmit the downlink beacon light a) with linearly polarised light and b) with a fixed polarisation base orientation with respect to that of the QKD channel. Therefore, detectors to determine the azimuth angle of the incoming polarisation state are implemented in both classical receiver channels of the QATS box.

An additional multimode fibre coupling stage is implemented in the SWIR band, allowing the QATS box to be used also in LEO-DTE scenarios. Downlink data rates of up to 2.5 Gbps can be achieved with the implemented fibre.

The modular design of the QATS box also supports a future upgrade with an adaptive optics system. The use of adaptive optics would allow single mode fibre coupling of the received downlink light in both the QKD and SWIR channels. The single mode fibre in the QKD channel would then act as a filter element with an extremely narrow field of view to reject background radiation.

\subsection{Motorized QKD module}

The fully motorized QKD module will be integrated within the QATS module as shown in Figure \ref{fig:fullsetup}. Most of the QKD satellite missions expected in the near future will use downlink discrete variable DV-QKD signal protocols, such as BB84 or BBM92, in which the encoding is based on the polarization degree of freedom of the photons. For this reason, the QKD module performs projective measurements of the polarization degree of freedom in two mutually unbiased bases by means of 4 free-space Silicon single photon counting detectors. 

During a satellite pass, the orientation of the satellite's frame of reference changes with respect to the ground station, leading to a dynamic polarization mismatch. The telescope and the rest of the optical path within the QATS also introduce non-negligible polarization transformations. Both contributions need to be compensated during a satellite pass, which can be done with a motorized set of two quarter- and one half-wave plates included in the QKD module. 
While the required corrections typically may be predicted from the trajectory of the satellite, one possible mode of operation is the correction with a pre-calculated pattern, i.e. an open loop control. Alternatively, if the downlink beacon light may be used as a polarization reference, a closed-loop operation mode can be realized with the help of the polarization rotation detectors available in both VIS and SWIR bands.

Furthermore, spatial and spectral filters are arranged to minimize the background noise. The QKD module includes motorised filter stages to suppress out-of-band background radiation. The default set of filters supports QKD wavelengths equal to $780 \pm 10$ nm and $850 \pm 3$ nm. These specific bands were chosen to be compatible with the majority of QKD satellite missions, while changing the filters can allow for operation at a different wavelength inside the band from 780 nm to 900~nm.

\subsection{Multi-band Uplink Beacon module}
\begin{figure}
    \centering
    \includegraphics[width=0.9\textwidth]{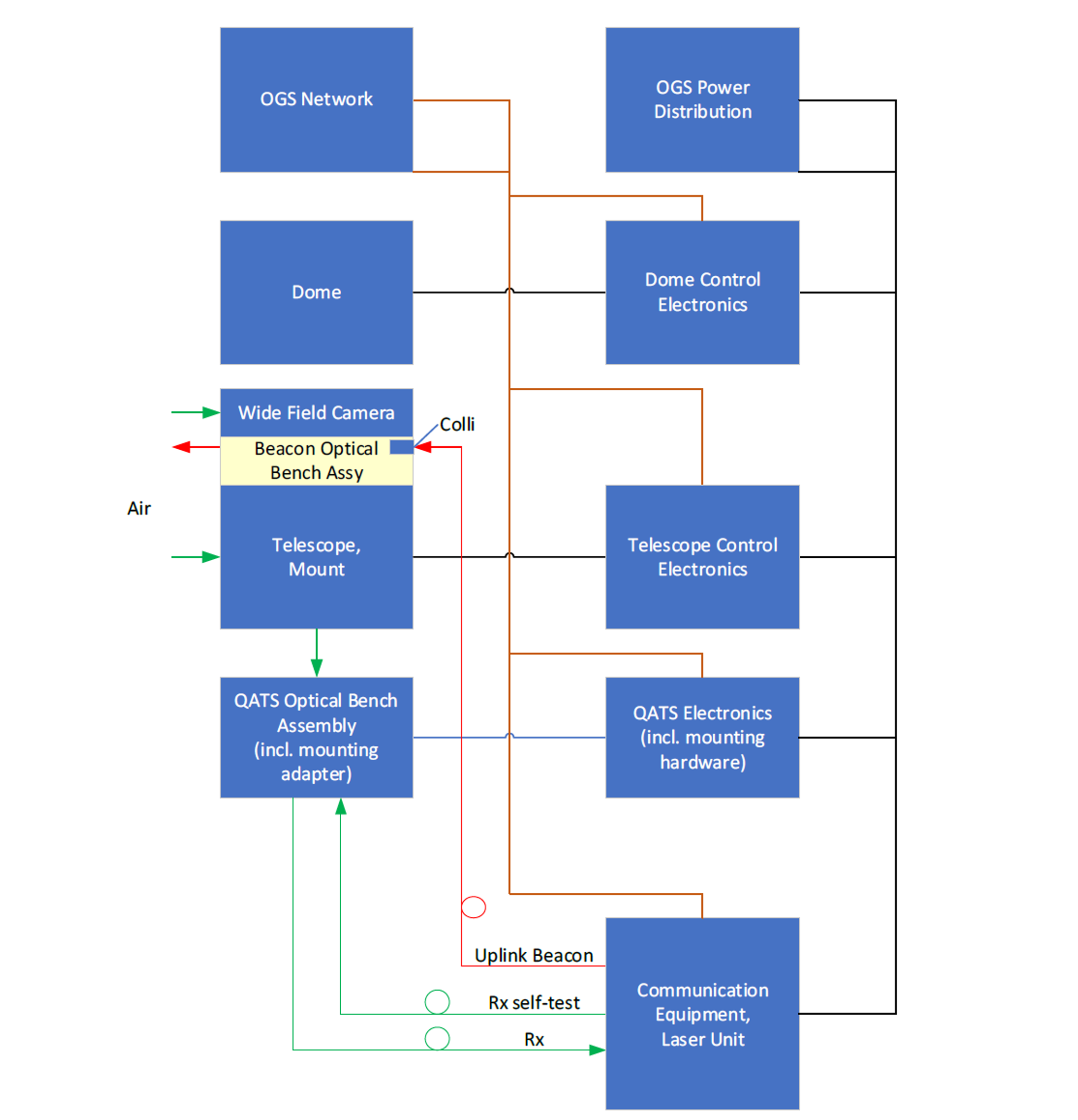}
    \caption{Top level functional ADQOGS block diagram.}
    \label{fig:toplevel}
\end{figure}

A simplified functional top-level block diagram of an optical ground station (ADQOGS) is shown in Figure \ref{fig:toplevel}. A Wide Field Camera will be used for e.g. initial coarse pointing model acquisition or large field of view observations. 
The BOBA is an OGS subsystem that is physically located at the telescope body, see Figure \ref{fig:fullsetup}. It enables ADQOGS to transmit a modulated high-power uplink beam towards a communication partner terminal hosted onboard a LEO satellite. The designed system provides a beacon power of up to 10W and a spectral range from 1530 nm to 1610 nm. 
The BOBA module is equipped with a Low-frequency Tip/tilt correction of transmitted beam line of sight, with possible point ahead angle corrections. The co-alignment between the Rx direction (QATS Optical Bench) and the Tx direction (BOBA) is done via simultaneous star imaging. 

Moreover, TII is developing a second uplink beacon module, designed to deliver a high data rate coherent optical feeder-link at 1064 nm for next-generation telecommunication satellites. 

\subsection{Software}

\begin{figure}
    \centering
    \includegraphics[width=1\textwidth]{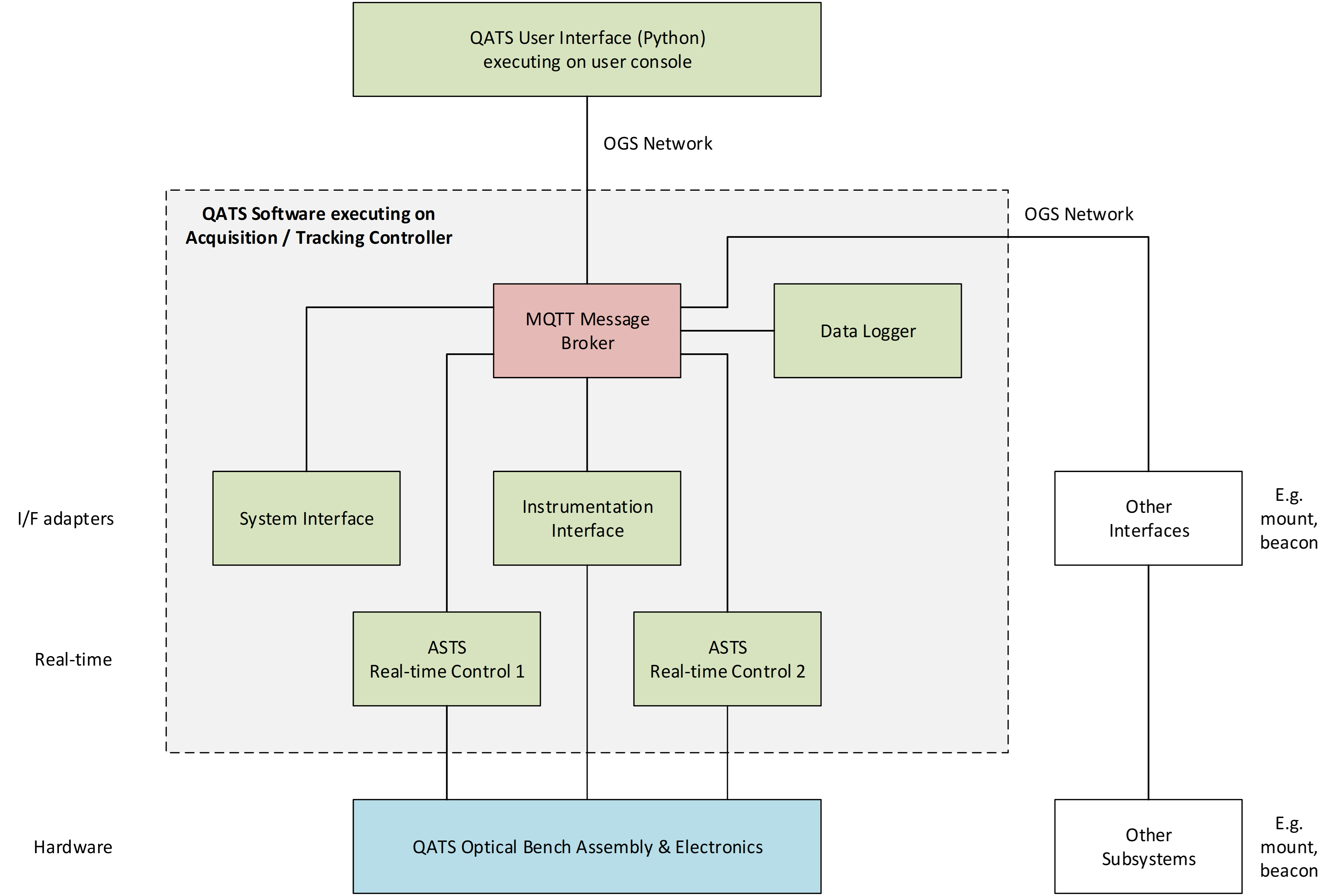}
    \caption{QATS Software Architecture.}
    \label{fig:software}
\end{figure}

Figure \ref{fig:software} shows the overall architecture of the software involved in QATS operation. The main components are:
\begin{itemize}
    \item QATS Software, executing on the Acquisition / Tracking Controller.
    \item QATS User Interface, executable on any computer connected to the OGS network.
\end{itemize}

The QATS Software and QATS User Interface make use of an off-the-shelf MQTT message broker, to connect all components via a common message bus. MQTT is a standard messaging protocol for distributed systems. It is designed as a lightweight publish–subscribe messaging transport that is ideal for connecting remote devices with a small code footprint and minimal network bandwidth.

\section{Preliminary testing results}
\label{sec:testing} 

\begin{figure}
    \centering
    \includegraphics[width=0.9\textwidth]{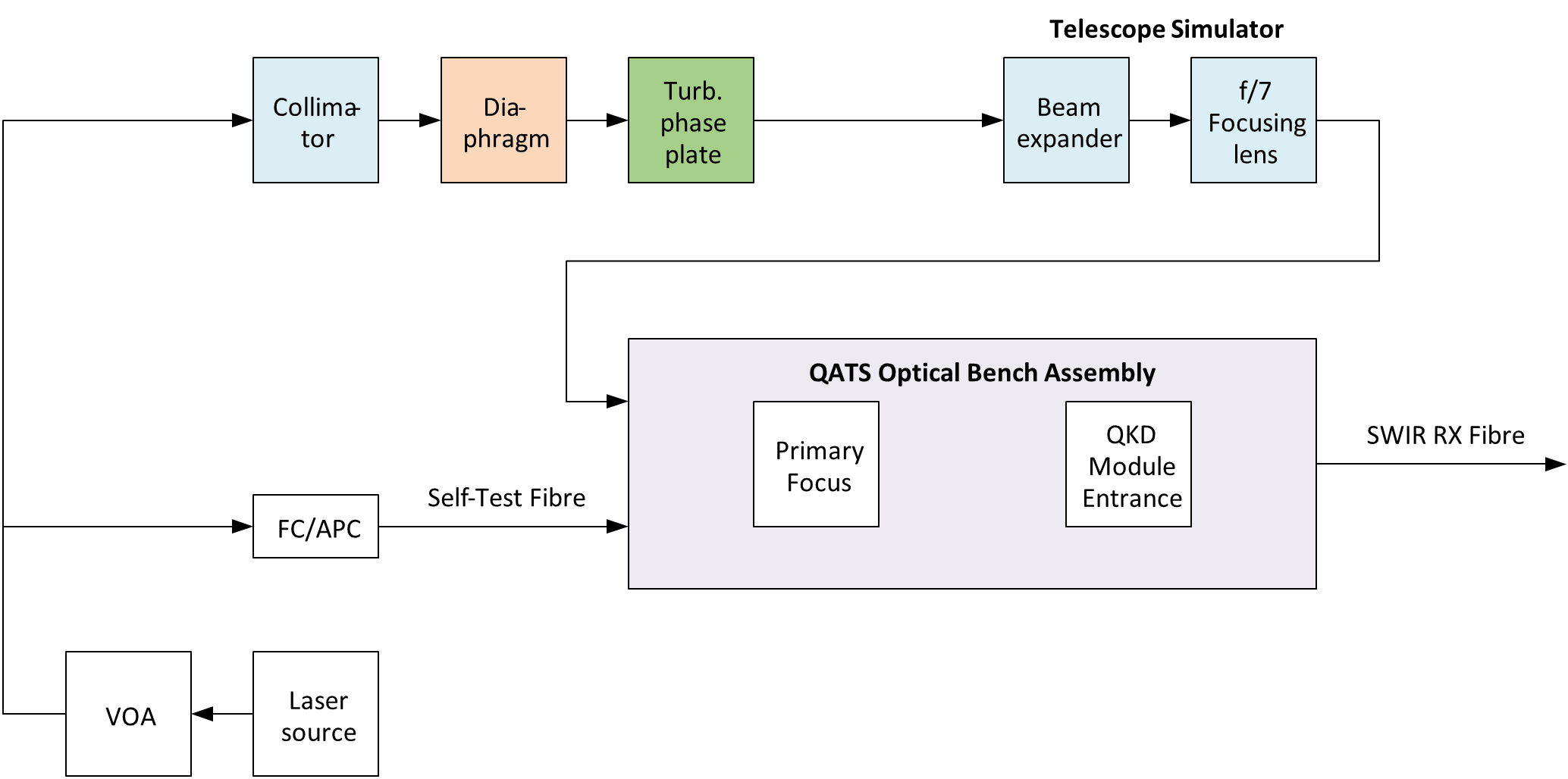}
    \caption{QATS Functionality test setup. VOA:  Variable Optical Attenuator. FC/APC: fiber connector.}
    \label{fig:FATsetup}
\end{figure}

The QATS functionality test is scheduled for October 2024 and split into three main activities:

\begin{itemize}
    \item Functional tests in the VIS band
    \item Optical tests in the QKD band
    \item Functional tests in the SWIR band
\end{itemize}
The sketch in Figure \ref{fig:FATsetup} depicts the overall test setup.

To simulate atmospheric turbulence effects in the functional tests, a rotating turbulence phase plate is inserted into the optical beam path of the test setup. With an effective magnification of 100 (800 mm telescope / 8 mm test beam), the simulated Fried parameters, at the line of sight and at 550 nm, are 46 mm and 18 mm respectively. With the turbulence generator boundary-layer wind speeds up to 11 m/s can be simulated.

\begin{figure}
    \centering
    \includegraphics[width=0.8\textwidth]{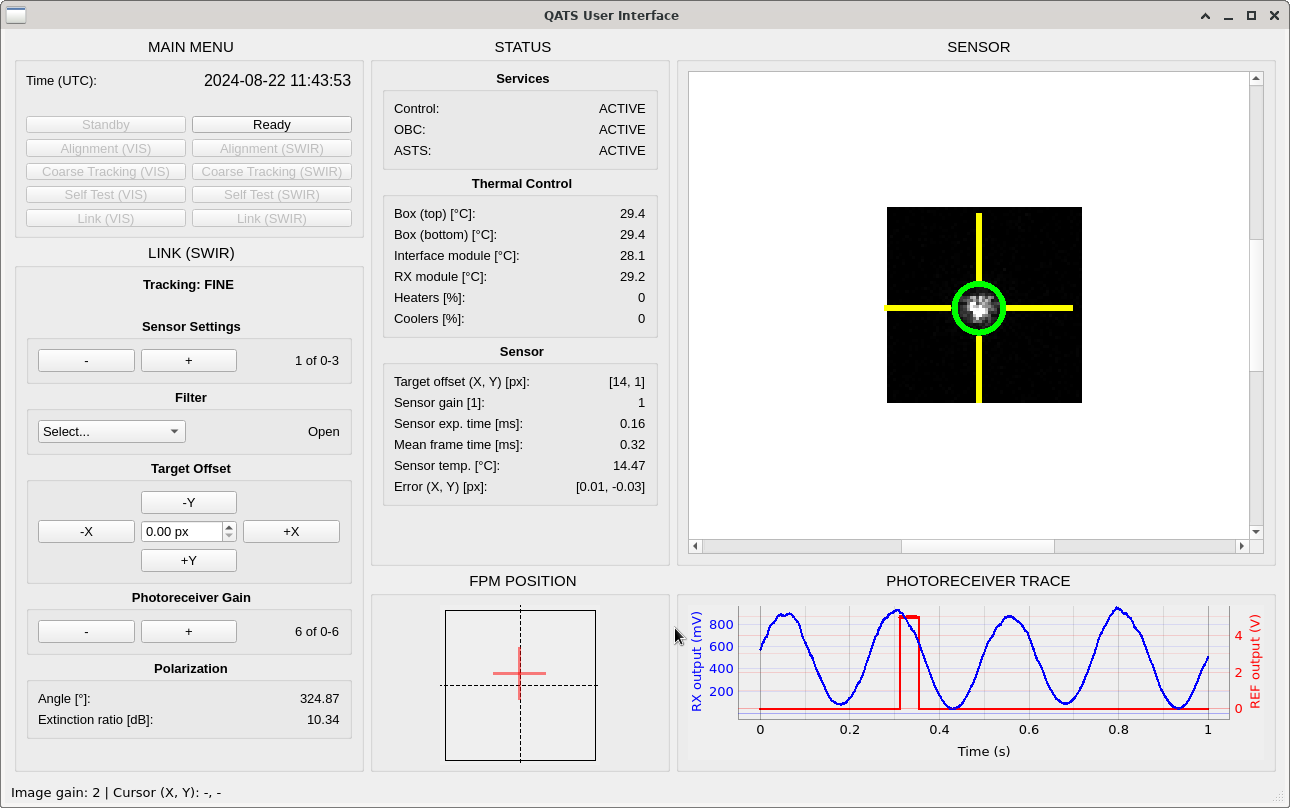}
    \caption{Screenshot of User Interface during tracking test with simulated strong atmospheric turbulence.}
    \label{fig:screenshot}
\end{figure}

A first preliminary test result is depicted in the screenshot Figure \ref{fig:screenshot}. In this test the tracking capability of the SWIR channel tracking system was examined. During the test the minimum Fried parameter and maximum lateral wind speed were simulated. The tracking system could lock without any problem also the polarization state azimuth angle measurement delivered satisfactory results.

\section{CONCLUSIONS}
\label{sec:conclusion}
In this article, we have presented the current status of the establishment of a versatile quantum reception optical ground station in Abu Dhabi. ADQOGS aims to become a trusted node for the global quantum network, offering compatibility with most of the upcoming QKD satellite missions. ADQOGS is open to diversifying its missions and could potentially be used for optical feeder links, space situational awareness, laser ranging, deep space communication, and more free-space optics related activities.




\begin{thebibliography}{1}

\bibitem{zhou2022quantum}
Zhou, H., Lv, K., Huang, L., and Ma, X., ``Quantum network: security assessment
  and key management,'' {\em IEEE/ACM Transactions on Networking}~{\bf 30}(3),
  1328--1339 (2022).

\bibitem{de2024parallel}
De~Santis, G., Kravtsov, K., Amairi-Pyka, S., and Grieve, J.~A., ``Parallel
  trusted node approach for satellite quantum key distribution,'' {\em arXiv
  preprint arXiv:2406.08562}  (2024).

\bibitem{fischer2021modular}
Fischer, E., Kudielka, K., Brady, A., Kamm, A., Berkefeld, T., and Ursin, R.,
  ``Modular adaptive optics solution for a qkd receiver on a fork mount
  telescope system,'' in [{\em International Conference on Space Optics—ICSO
  2020}{\nolinebreak\hspace{0.1em}]},   {\bf 11852},  380--387, SPIE (2021).

\end{thebibliography}
\end{document}